\documentstyle[12pt]{article} 
\catcode`\@=11

\def\psfortextures{
\def\PSspeci@l##1##2{%
\special{illustration ##1\space scaled ##2}}}

\def\psfordvips{
\def\PSspeci@l##1##2{%
\d@my=0.1bp \d@mx=\drawingwd \divide\d@mx by\d@my%
\includegraphics{##1\space}}}

\def\psforoztex{
\def\PSspeci@l##1##2{%
\special{##1 \space
      ##2 1000 div dup scale
      \putsp@ce{\number-\psllx} \putsp@ce{\number-\pslly} translate}}}
\def\putsp@ce#1{#1 }

\def\psonlyboxes{
\def\PSspeci@l##1##2{%
\at{0cm}{0cm}{\boxit{\vbox to\drawinght
  {\vss
  \hbox to\drawingwd{\at{0cm}{0cm}{\hbox{(##1)}}\hss}
  }}}}}

\newdimen\drawinght\newdimen\drawingwd
\newdimen\psxoffset\newdimen\psyoffset
\newbox\drawingBox

\newread\epsffilein    
\newif\ifepsffileok    
\newif\ifepsfbbfound   
\newif\ifepsfverbose   
\newdimen\epsfxsize    
\newdimen\epsfysize    
\newdimen\epsftsize    
\newdimen\epsfrsize    
\newdimen\epsftmp      
\newdimen\pspoints     
\def\ReadPSize#1{
\edef\PSfilename{#1}
\global\def\epsfllx{72}
\global\def\epsflly{72}
\global\def\epsfurx{540}
\global\def\epsfury{720}
\openin\epsffilein=#1
\ifeof\epsffilein\errmessage{I couldn't open #1, will ignore it}\else
   {\epsffileoktrue \chardef\other=12
    \def\do##1{\catcode`##1=\other}\dospecials \catcode`\ =10
    \loop
       \read\epsffilein to \epsffileline
       \ifeof\epsffilein\epsffileokfalse\else
          \expandafter\epsfaux\epsffileline:. \\%
       \fi
   \ifepsffileok\repeat
   \ifepsfbbfound\else
    \ifepsfverbose\message{No bounding box comment in #1;
using defaults}\fi\fi
   }\closein\epsffilein\fi
\def\psllx{\epsfllx}\def\pslly{\epsflly}%
\def\psurx{\epsfurx}\def\psury{\epsfury}%
\drawinght=\epsfury bp%
\advance\drawinght by-\epsflly bp%
\drawingwd=\epsfurx bp%
\advance\drawingwd by-\epsfllx bp%
}

{\catcode`\%=12 \global\let\epsfpercent=
\long\def\epsfaux#1#2:#3\\{\ifx#1\epsfpercent
   \def\testit{#2}\ifx\testit\epsfbblit
      \epsfgrab #3 @ @ @ \\%
      \epsffileokfalse
      \global\epsfbbfoundtrue
   \fi\else\ifx#1\par\else\epsffileokfalse\fi\fi}%

\def\epsfgrab #1 #2 #3 #4 #5\\{%
   \global\def\epsfllx{#1}\ifx\epsfllx\empty
      \epsfgrab #2 #3 #4 #5 @\\\else
   \global\def\epsflly{#2}%
   \global\def\epsfurx{#3}\global\def\epsfury{#4}\fi}%

\newcount\xscale \newcount\yscale \newdimen\pscm\pscm=1cm
\newdimen\d@mx \newdimen\d@my
\let\ps@nnotation=\relax
\def\psboxto(#1;#2)#3{\vbox{
   \catcode`\:=12
   \ReadPSize{#3}
   \divide\drawingwd by 1000
   \divide\drawinght by 1000
   \d@mx=#1
   \ifdim\d@mx=0pt\xscale=1000
         \else \xscale=\d@mx \divide \xscale by \drawingwd\fi
   \d@my=#2
   \ifdim\d@my=0pt\yscale=1000
         \else \yscale=\d@my \divide \yscale by \drawinght\fi
   \ifnum\yscale=1000
         \else\ifnum\xscale=1000\xscale=\yscale
                    \else\ifnum\yscale<\xscale\xscale=\yscale\fi
              \fi
   \fi
   \divide \psxoffset by 1000\multiply\psxoffset by \xscale
   \divide \psyoffset by 1000\multiply\psyoffset by \xscale
   \global\divide\pscm by 1000
   \global\multiply\pscm by\xscale
   \multiply\drawingwd by\xscale \multiply\drawinght by\xscale
   \ifdim\d@mx=0pt\d@mx=\drawingwd\fi
   \ifdim\d@my=0pt\d@my=\drawinght\fi
\message{[#3\space [BoundingBox\string:
\space\epsfllx\space\epsflly\space\epsfurx\space\epsfury]}%
\message{[scaled\space\the\xscale\string:
\space\the\drawingwd\space x \the\drawinght]]}%
 \hbox to\d@mx{\hss\vbox to\d@my{\vss
   \global\setbox\drawingBox=\hbox to 0pt{\kern\psxoffset\vbox to 0pt{
      \kern-\psyoffset
      \PSspeci@l{\PSfilename}{\the\xscale}
      \vss}\hss\ps@nnotation}
   \global\ht\drawingBox=\the\drawinght
   \global\wd\drawingBox=\the\drawingwd
   \baselineskip=0pt
   \copy\drawingBox
 \vss}\hss}
  \global\psxoffset=0pt
  \global\psyoffset=0pt
  \global\pscm=1cm
  \global\drawingwd=\drawingwd
  \global\drawinght=\drawinght
}}

\def\psboxscaled#1#2{\vbox{
  \catcode`\:=12
  \ReadPSize{#2}
  \xscale=#1
  \divide\drawingwd by 1000\multiply\drawingwd by\xscale
  \divide\drawinght by 1000\multiply\drawinght by\xscale
  \divide \psxoffset by 1000\multiply\psxoffset by \xscale
  \divide \psyoffset by 1000\multiply\psyoffset by \xscale
  \global\divide\pscm by 1000
  \global\multiply\pscm by\xscale
\message{[#2\space [BoundingBox\string:
\space\epsfllx\space\epsflly\space\epsfurx\space\epsfury]}%
\message{[scaled\space\the\xscale\string:
\space\the\drawingwd\space x \the\drawinght]]}%
  \global\setbox\drawingBox=\hbox to 0pt{\kern\psxoffset\vbox to 0pt{
     \kern-\psyoffset
     \PSspeci@l{\PSfilename}{\the\xscale}
     \vss}\hss\ps@nnotation}
  \global\ht\drawingBox=\the\drawinght
  \global\wd\drawingBox=\the\drawingwd
  \baselineskip=0pt
  \copy\drawingBox
  \global\psxoffset=0pt
  \global\psyoffset=0pt
  \global\pscm=1cm
  \global\drawingwd=\drawingwd
  \global\drawinght=\drawinght
}}

\def\psannotate#1#2{\def\ps@nnotation{#2\global\let\ps@nnotation=\relax}#1}
\def\pscaption#1#2{\vbox{
   \setbox\drawingBox=#1
   \copy\drawingBox
   \vskip\baselineskip
   \vbox{\hsize=\wd\drawingBox\setbox0=\hbox{#2}
     \ifdim\wd0>\hsize
       \noindent\unhbox0\tolerance=5000
    \else\centerline{\box0}
    \fi
}}}

\def\at#1#2#3{\setbox0=\hbox{#3}\ht0=0pt\dp0=0pt
  \rlap{\kern#1\vbox to0pt{\kern-#2\box0\vss}}}

\newdimen\gridht \newdimen\gridwd
\def\gridfill(#1;#2){
  \setbox0=\hbox to 1\pscm
  {\vrule height1\pscm width.4pt\leaders\hrule\hfill}
  \gridht=#1
  \divide\gridht by \ht0
  \multiply\gridht by \ht0
  \advance \gridht by \ht0
  \gridwd=#2
  \divide\gridwd by \wd0
  \multiply\gridwd by \wd0
  \advance \gridwd by \wd0
  \vbox to \gridht{\leaders\hbox to\gridwd{\leaders\box0\hfill}\vfill}}

\def\frameit#1#2#3{\hbox{\vrule width#1\vbox{
  \hrule height#1\vskip#2\hbox{\hskip#2\vbox{#3}\hskip#2}%
        \vskip#2\hrule height#1}\vrule width#1}}
\def\boxit#1{\frameit{0.4pt}{0pt}{#1}}

\catcode`\@=12 

\psfordvips

\jot = 1.5ex
\def\baselinestretch{1.65}
\parskip 5pt plus 1pt

\catcode`\@=11

\@addtoreset{equation}{section}

\def\@normalsize{\@setsize\normalsize{15pt}\xiipt\@xiipt
\abovedisplayskip 14pt plus3pt minus3pt%
\belowdisplayskip \abovedisplayskip
\abovedisplayshortskip  \z@ plus3pt%
\belowdisplayshortskip  7pt plus3.5pt minus0pt}
\def\small{\@setsize\small{13.6pt}\xipt\@xipt
\abovedisplayskip 13pt plus3pt minus3pt%
\belowdisplayskip \abovedisplayskip
\abovedisplayshortskip  \z@ plus3pt%
\belowdisplayshortskip  7pt plus3.5pt minus0pt
\def\@listi{\parsep 4.5pt plus 2pt minus 1pt
            \itemsep \parsep
            \topsep 9pt plus 3pt minus 3pt}}

\def\underline#1{\relax\ifmmode\@@underline#1\else
        $\@@underline{\hbox{#1}}$\relax\fi}
\@twosidetrue
\relax

\catcode`@=12

\evensidemargin 0.0in
\oddsidemargin 0.0in
\topmargin -0.2in
\textwidth 6.4in
\textheight 8.9in

\catcode`\@=11

\def\section{\@startsection{section}{1}{\z@}{3.5ex plus 1ex minus
   .2ex}{2.3ex plus .2ex}{\large\bf}}

\def\ps@headings{\def\@oddfoot{}\def\@evenfoot{}
\def\@oddhead{\hbox{}\hfill
        \makebox[.5\textwidth]{\raggedright\ignorespaces --\thepage{}--
        \hfill }}
\def\@evenhead{\@oddhead}
\def\subsectionmark##1{\markboth{##1}{}}
}

\ps@headings

\catcode`\@=12

\relax

\def\figcap{\section*{Figure Captions\markboth
        {FIGURECAPTIONS}{FIGURECAPTIONS}}\list
        {Fig. \arabic{enumi}:\hfill}{\settowidth\labelwidth{Fig. 999:}
        \leftmargin\labelwidth
        \advance\leftmargin\labelsep\usecounter{enumi}}}
 \relax
\def\tablecap{\section*{Table Captions\markboth
        {TABLECAPTIONS}{TABLECAPTIONS}}\list
        {Table \arabic{enumi}:\hfill}{\settowidth\labelwidth{Table 999:}
        \leftmargin\labelwidth
        \advance\leftmargin\labelsep\usecounter{enumi}}}
 \relax
\def\reflist{\section*{References\markboth
        {REFLIST}{REFLIST}}\list
        {[\arabic{enumi}]\hfill}{\settowidth\labelwidth{[999]}
        \leftmargin\labelwidth
        \advance\leftmargin\labelsep\usecounter{enumi}}}
 \relax

\catcode`\@=11

\def\marginnote#1{}

\def\ps@headings{\def\@oddfoot{}\def\@evenfoot{}
\def\@oddhead{\hbox{}\hfill
        \makebox[.5\textwidth]{\raggedright\ignorespaces --\thepage{}--
        \hfill }}
\def\@evenhead{\@oddhead}
\def\subsectionmark##1{\markboth{##1}{}}
}

\ps@headings

\relax

\def\firstpage#1#2#3#4#5#6{
\begin{document}
\begin{titlepage}
\nopagebreak
\title{\begin{flushright}
        \vspace*{-1.0in}
     {\normalsize NUB--#1 #2}\\[-9mm]
       {\normalsize hep-th/0106244}\\[14mm]
\end{flushright}
{#3}}
\author{\large #4 \\ #5}
\maketitle
\vskip -7mm
\nopagebreak
\def\baselinestretch{1.0}
\begin{abstract}
{\noindent #6}
\end{abstract}
\vfill
\begin{flushleft}
\rule{16.1cm}{0.2mm}\\[-3mm]
$^{\star}${\small Research supported in part by
the National Science Foundation under grant
PHY--99--01057.}\\
June 2001
\end{flushleft}
\thispagestyle{empty}
\end{titlepage}}
\newcommand{\Zint}{{\mbox{\sf Z\hspace{-3.2mm} Z}}}
\newcommand{\Real}{{\mbox{I\hspace{-2.2mm} R}}}
\def\simlt{\stackrel{<}{{}_\sim}}
\def\simgt{\stackrel{>}{{}_\sim}}
\newcommand{\dal}{\raisebox{0.085cm}
{\fbox{\rule{0cm}{0.07cm}\,}}}
\newcommand{\dt}{\partial_{\langle T\rangle}}
\newcommand{\dtbar}{\partial_{\langle\bar{T}\rangle}}
\newcommand{\al}{\alpha^{\prime}}
\newcommand{\mst}{M_{\scriptscriptstyle \!S}}
\newcommand{\mpl}{M_{\scriptscriptstyle \!P}}
\newcommand{\dv}{\int{\rm d}^4x\sqrt{g}}
\newcommand{\lv}{\left\langle}
\newcommand{\rv}{\right\rangle}
\newcommand{\ph}{\alpha}
\newcommand{\abar}{\bar{a}}
\newcommand{\sbar}{\,\bar{\! S}}
\newcommand{\xbar}{\,\bar{\! X}}
\newcommand{\fbar}{\,\bar{\! F}}
\newcommand{\zbar}{\bar{z}}
\newcommand{\dbar}{\,\bar{\!\partial}}
\newcommand{\tbar}{\bar{T}}
\newcommand{\taubar}{\bar{\tau}}
\newcommand{\ubar}{\bar{U}}
\newcommand{\tetabar}{\bar\tau}
\newcommand{\etabar}{\bar\eta}
\newcommand{\qbar}{\bar q}
\newcommand{\ybar}{\bar{Y}}
\newcommand{\phb}{\bar{\alpha}}
\newcommand{\cm}{Commun.\ Math.\ Phys.~}
\newcommand{\prl}{Phys.\ Rev.\ Lett.~}
\newcommand{\pr}{Phys.\ Rev.\ D~}
\newcommand{\pl}{Phys.\ Lett.\ B~}
\newcommand{\ibar}{\bar{\imath}}
\newcommand{\jbar}{\bar{\jmath}}
\newcommand{\np}{Nucl.\ Phys.\ B~}
\newcommand{\F}{{\cal F}}
\renewcommand{\L}{{\cal L}}
\newcommand{\A}{{\cal A}}
\newcommand{\M}{{\cal M}}
\newcommand{\N}{{\cal N}}
\newcommand{\T}{{\cal T}}
\newcommand{\ads}{{\rm AdS}}
\renewcommand{\Im}{\mbox{Im}}
\newcommand{\e}{{\rm e}}
\newcommand{\be}{\begin{equation}}
\newcommand{\en}{\end{equation}}
\newcommand{\gsi}{\,\raisebox{-0.13cm}{$\stackrel{\textstyle
>}{\textstyle\sim}$}\,}
\newcommand{\lsi}{\,\raisebox{-0.13cm}{$\stackrel{\textstyle
<}{\textstyle\sim}$}\,}
\date{}
\firstpage{3216}{}
{\large\sc Renormalization of Boundary Fermions  and\\[-5mm]
World-Volume Potentials on D-branes$^\star$} {P. Khorsand and T.R.
Taylor} {\normalsize\sl Department of Physics, Northeastern
University, Boston, MA 02115, U.S.A.} {We consider a sigma model
formulation of open string theory with boundary fermions carrying
Chan-Paton charges at the string ends. This formalism is
particularly \linebreak suitable for studying world-volume potentials on
D-branes. We perform explicit two-loop computations 
of the potential  T-dual to the non-abelian Born-Infeld action.
We also discuss the world-volume couplings of NS fluxes which are
responsible for Myers' dielectric effect.} \setcounter{section}{0}
\section{Introduction}
Open strings end on D-branes. At low energies, their
gauge degrees of freedom are described by
a non-linear generalization
of Maxwell's electrodynamics -- the Born-Infeld theory on the D-brane
world-volume. When a number $N$ of D-branes occupy the same hypersurface,
the gauge group
gets enhanced to $U(N)$. The world-volume action of $N$ coincident D-branes
is described then by a non-abelian generalization of Born-Infeld theory (NBI).
In spite of many efforts to unveil the structure of NBI action,
its detailed form still remains elusive \cite{revs}.

There are several methods available for studying D-brane actions.
One of the most effective ones
relies on the computations of scattering amplitudes for $U(N)$ gauge
bosons. In fact, most results and conjectures are based on
the higher-derivative corrections to four-point amplitudes analyzed 
by Tseytlin
\cite {tsey} in order to extract all interaction
terms up to the fourth order in the gauge
field strength
$F$. Unfortunately, it is very difficult to push this type of computations
any further
since they involve amplitudes with larger numbers
of external gauge bosons, a very difficult task indeed until a good
recursive technique becomes available. Another method is based
on world-sheet sigma model computations. Here, the world-volume action
is reconstructed from the equations of motion which are obtained by
requiring conformal invariance on the world-sheet. 
However, these computations \cite{dorn,perry} appear to be quite
complicated. There has not been much work done in this direction
over the past few years.

The reason why the sigma model approach becomes difficult in the presence
of non-abelian gauge symmetry is the non-linear (Wilson loop) coupling
of the open string boundary $\partial\Sigma$ 
to the gauge field background.
For the bosonic string, the relevant part of the
sigma model action for string embeddings
$X^{\mu}(z,\bar{z})$ coupled to the background gauge fields $A_{\mu}(X)$ is
\be
S={1\over 2\pi\al}\int_{\Sigma}d^2z\,\partial X^{\mu}\bar{\partial}X_{\mu}
+S_{\partial\Sigma}(A).\label{bos}\en
The boundary coupling
is given by the Wilson loop:
\be e^{-S_{\partial\Sigma}(A)}=\makebox{Tr}\,{\cal P}\,
e^{i\oint_{\partial\Sigma}
d\tau A_{\mu}(X)\partial_{\tau}X^{\mu}},\label{wloop} \en
where  $A_{\mu}(X)$ are Hermitian matrices in the fundamental
representation of the gauge group algebra and $\cal P$ is the path-ordering
operation. 
Here, $\partial_{\tau}$ denotes the tangential derivative.
In the traditional sigma model approach, one splits the
embeddings into classical and quantum parts, expands the gauge field
background $A_{\mu}(X)$
around the classical part and integrates over 
the quantum  string fluctuations.
The classical field is treated then as a sigma model coupling
and the requirement that the corresponding beta function be zero
is interpreted as its equation of motion. However, this procedure
involves the Wilson loop expansion
and leads to complicated Feynman diagrams.

In the case of D-branes of codimension $d> 0$, the sigma model
background includes also the scalar  fields $\phi_i$, $i=1,\dots,
d$, in the adjoint representation of the gauge group. They
describe  brane oscillations in the directions transverse to the
rigid hypersurface. These fields are associated with the
world-sheet coordinates $X^i$ satisfying the Dirichlet boundary
conditions, $X^i(\partial\Sigma)=0$, and their boundary coupling
is \be e^{-S_{\partial\Sigma}(\phi)}=\makebox{Tr}\,{\cal P}\,
e^{i\oint_{\partial\Sigma} d\tau\phi_i(X)\partial_rX^i}, \en 
where $\partial_r$ denotes the normal derivative. The
main focus of our work is the effective potential for these
non-abelian scalars. In fact, the scalar lagrangian is related to
the gauge lagrangian by T-duality and can be used to extract the
NBI action.

This paper is organized as follows. In Section 2, we reformulate the sigma
model by introducing boundary fermions that carry Chan-Paton charges at the
string
ends. In Section 3, we describe the one-loop computation of anomalous
dimensions of these fermions 
and of the beta function associated to non-abelian scalar couplings.
We extend our discussion to two loops for bosonic strings and superstrings
in Sections 4 and 5, respectively. In particular, we derive the world-volume
D-brane potentials and discuss the relation of our results to
previous sigma model computations of the NBI action.
In Section 6, we study the effects of background NS fluxes up to two
loops in the sigma model expansion. At one loop, we recover Myers' dielectric
coupling \cite{myers} in both bosonic and in superstring cases. 
We also show that this coupling does
not receive any two-loop corrections.

\section{Sigma model with boundary fermions}
It is well known that gauge degrees of freedom can be introduced in open string
theory by attaching to the string boundary a set of complex Grassmann valued
fields  $\lambda_a$,
$a=1,\dots, N $, in the fundamental representation of $U(N)$.
Quantization of these fermions generates $N$ conserved charges
attached to the string boundary and generates Chan-Paton factors.
In the context of sigma model, the Wilson loop action can be
rewritten \cite{dhok} as
a functional integral over the boundary fermions coupled to the string
coordinates $X^i$ via $N{\times}N$ Hermitian, traceless\footnote{The 
traces describe the center of mass motion which
decouples  from other degrees of freedom.} matrices $\phi_i$:
\be e^{-S_{\partial\Sigma}(\phi)}=\sum_{\theta}\int_{\makebox{\scriptsize AP}}
[d\lambda^{\dagger}
d\lambda]\,e^{i\theta[\lambda^{\dagger}\lambda(\tau=\tau_0)+ {N\over 2}-1]}\;
e^{-\int_{0}^{2\pi}d\tau
(\lambda^{\dagger}{d\over\partial\tau}\lambda
-i\partial_rX^i\lambda^{\dagger}\phi_i\lambda)}.\label{dhok}
\en
The fermions satisfy antiperiodic (AP) boundary conditions on the interval
$\tau\in[0,2\pi]$. The summation, which extends over the angles
$\theta=2\pi k/N$ with $k=1,\dots, N$, must be included in order to project
the intermediate states in the path integral on coherent states of occupation
number 1. The point $\tau_0$ can be arbitrarily chosen on the boundary.
In this work, we study the renormalization
properties of the sigma model reformulated by using these boundary
fermions.\footnote{See \cite{fermions} for a partial list of references
utilizing similar ``auxiliary'' fermions.}
We limit our
considerations to the case of constant, spacetime-independent
backgrounds $\phi_i$.

We first consider quantum effects due to fluctuations of string
embeddings $X^i$. Integration over these fluctuations
leads to  ultraviolet divergences. We focus on the loop
corrections to the correlation functions that determine the beta
function $\beta_i(\phi)$. Since the classical theory is conformally 
invariant, we are free to work on a disk, parameterized by $(r,\tau)$,
with the
boundary at $r=1$. The bosonic string fluctuations couple to the
boundary through their derivatives $\partial_rX^i$, therefore the
corresponding Feynman diagrams involve the two-point functions
\be\langle\partial_rX^i(\tau)\partial_{r'}X^j(\tau')\rangle\equiv
\left.\langle\partial_rX^i(r,\tau)\partial_{r'}X^j(r',\tau')
\rangle\right|_{r=r'=1}=\delta^{ij}{\partial\over\partial r}
\left.{\partial\over\partial r'}{\cal D}(r,\tau;r',\tau')
\right|_{r=r'=1},\en where $\cal D$ is the Dirichlet propagator.
The r.h.s.\ is related to the Neumann propagator $\cal N$ by a
T-duality relation, \be \left.{\partial\over\partial
r}{\partial\over\partial r'} {\cal
D}(r,\tau;r',\tau')\right|_{r=r'=1}= -\left.{\partial\over\partial
\tau}{\partial\over\partial \tau'} {\cal
N}(r,\tau;r',\tau')\right|_{r=r'=1}={\al\over 2\sin^{2}[{\scriptstyle{
(\tau-\tau')\over 2}}]} .\label{trel} \en
It is convenient to define
\be D(\tau-\tau')\equiv {\al\over 2\sin^{2}[{\scriptstyle{
(\tau-\tau')\over 2}}]},\en
so that
\be\langle\partial_rX^i(\tau)\partial_{r'}X^j(\tau')
\rangle= D(\tau-\tau') \delta^{ij}\, .\en 

The
sigma model contains only one type of interaction, \be
S_I=i\int_{0}^{2\pi}\!\!d\tau\,\partial_rX^i
\lambda^{\dagger}_a(\phi_i)^{ab}\lambda_b\, ,\label{iterm} \en
which is essentially a QED-like vertex with $d$ $N{\times}N$
coupling matrices $\phi_i$. 

Finally, the free AP fermionic
propagator is \be
\Delta_{ab}^{(0)}(\tau,\tau')=
\langle\lambda_a(\tau)\lambda^{\dagger}_b(\tau')
\rangle ={i\over 2}\delta_{ab}\,\makebox{sign}(\tau-\tau'). \en
\newpage\section{One-loop computations}
\begin{figure}
\[
\psannotate{\psboxto(0cm;5cm){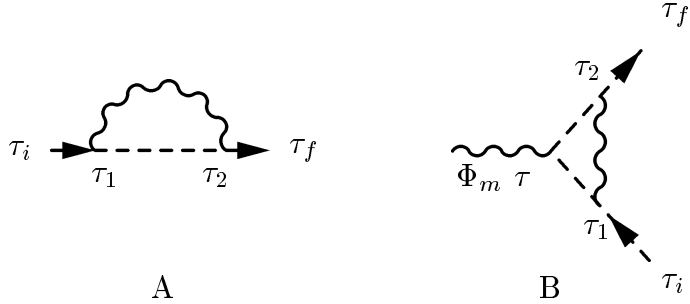}}{}
 \]\vskip -1cm
\caption{\em One-loop correction to the boundary fermion propagator {\rm (A)} 
and to the boundary fermion-string vertex {\rm (B)}.
Bosonic string coordinates are represented by wiggly lines while boundary 
fermions by dashed lines.}
\end{figure}

Although the one loop computation is very simple, it is quite
instructive to outline some details. 
\begin{figure}
\[
\psannotate{\psboxto(0cm;5cm){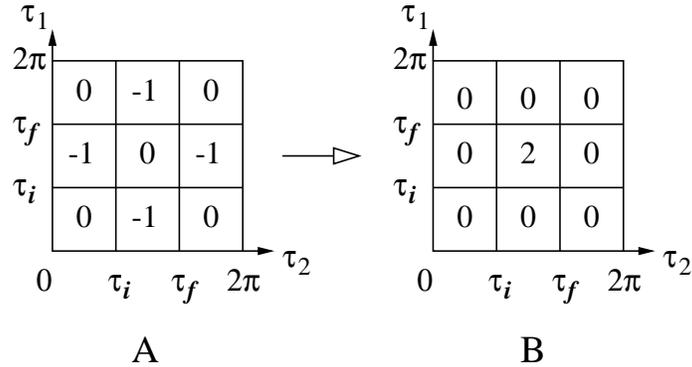}}{}
 \]\vskip -1cm
\caption{\em Weights of contributions to the one-loop propagator
from all integration regions, before {\rm (A)} and after {\rm (B)} 
the rearrangement that leads to path-ordering.}
\vskip 5mm
\end{figure}\nopagebreak
We begin with the one-loop
correction to the fermion propagator shown on Fig.1A:
\begin{eqnarray} {i\over 2}\makebox{sign}(\tau_i-\tau_f)
\Delta^{(1)}_{ab}(\tau_i,\tau_f)&=&\nonumber\\ & &\hskip -4cm{i\over 8}
(\phi_i\phi_i)_{ab}\int_{0}^{2\pi}\!\!\!\!d\tau_1\!\int_{0}^{2\pi}
\!\!\!d\tau_2\, \makebox{sign}(\tau_i-\tau_1)D(\tau_1-\tau_2)
\makebox{sign}(\tau_1-\tau_2) \makebox{sign}(\tau_2-\tau_f). \end{eqnarray}
Note that the integration over the interaction points $\tau_1$
and $\tau_2$ extends {\em a priori} over the whole range of angles
from 0 to $2\pi$, inside and outside the interval
$[\tau_i,\tau_f]$, with different integration regions
contributing $\int d\tau_1\int d\tau_2D(\tau_1-\tau_2)$ with $\pm$ 
signs depending on the relative position of the angles.
We display the relevant
signs on Fig.2A. 
We also show zeroes for regions that cancel as a
consequence of $\tau_1\leftrightarrow\tau_2$ symmetry of the
$D$-propagator. Since according to the relation (\ref{trel}), the
$D$-propagator is a double derivative of a periodic
function: \be D(\tau-\tau')=
-{\partial^2{\cal N}\over\partial \tau\partial \tau'}\; ,\qquad
 {\cal N}(\tau-\tau')=-\al\ln
\sin^{2}[{\scriptstyle{(\tau-\tau')\over 2}}]\; ,\en the
sum of contributions in each row or column of Fig.2A is zero. This
fact can be used to rearrange the integration regions as shown in
Fig.2B. Finally, by using the $\tau_1\leftrightarrow\tau_2$
symmetry one obtains 
\be \Delta^{(1)}_{ab}(\tau_i-\tau_f)=
(\phi_i\phi_i)_{ab}\;{\cal P}\! \int_{\tau_i}^{\tau_f} d\tau_1
d\tau_2D(\tau_1-\tau_2)\, ,\label{int} 
\en so the vertices become
path-ordered. As a result, the perturbation theory becomes
path-ordered, like in ordinary quantum mechanics. In the sigma
model framework, this amounts to Chan-Paton rules for inserting
the interaction vertex (\ref{iterm}).

The integral (\ref{int}) is divergent. In fact, it contains two
types of singularities: linear divergence along the line
$\tau_1=\tau_2$ and logarithmic ``end-point'' singularities. The
linear divergence is spurious and can be avoided by a formal
integration: \be {\cal P}\! \int_{\tau_i}^{\tau_f} d\tau_1
d\tau_2D(\tau_1-\tau_2)={1\over 2}\int_{\tau_i}^{\tau_f} d\tau_1
\int_{\tau_i}^{\tau_f}d\tau_2D(\tau_1-\tau_2) ={\cal
N}(\tau_i-\tau_f)-{\cal N}(0)\,. \en The remaining logarithmic singularity
${\cal N}(0)$ can be regulated by introducing a short-distance
cutoff $\Lambda$. As a result, one obtains \be
\Delta^{(1)}_{ab}(\tau_i-\tau_f)=\al (\phi_i\phi_i)_{ab}\; \ln\left(
{\Lambda\over\sin^{2}[{\scriptstyle{ (\tau_i-\tau_f)\over
2}}]}\right) \en The corresponding one-loop anomalous
dimension matrix of boundary fermions is \be
\gamma^{(1)}=\al\phi_i\phi_i\, . \label{one}\en

The one-loop correction to the interaction term (\ref{iterm}) is
shown on Fig.1B. For our purposes, it is sufficient to consider the
one-particle irreducible three-point function
$\Phi_m(\tau_i,\tau,\tau_f)$, with the tree-level coupling
$\phi_m$ inserted at  $\tau\in [\tau_i,\tau_f]$. By rearranging
integrals in a manner similar to the fermion propagator, one can
rewrite this diagram as a path-ordered integral, with the
position of interaction points correlated with the order of
non-abelian couplings in a way dictated by Chan-Paton rules: \be
\Phi_m^{(1)}(\tau_i,\tau,\tau_f)=\phi_i\phi_m\phi_i
\int_{\tau_i}^{\tau} d\tau_1\int_{\tau}^{\tau_f} d\tau_2\,
D(\tau_1-\tau_2)\, . \en The regularized integral yields \be
\Phi_m^{(1)}(\tau_i,\tau,\tau_f)=\al\phi_i\phi_m\phi_i \ln\left(
{\sin^{2}[{\scriptstyle{ (\tau-\tau_i)\over
2}}]\sin^{2}[{\scriptstyle{ (\tau-\tau_f)\over
2}}]\over\Lambda\sin^{2}[{\scriptstyle{ (\tau_i-\tau_f)\over
2}}]}\right). \en 

After combining this result with Eq.(\ref{one}), we find
the one-loop beta function \be
\beta_m^{(1)}=\al[\phi_i,[\phi_i,\phi_m]]\, .\label{phi} \en
The requirement of vanishing  one-loop beta function, {\em i.e}.\
the non-renormalization
of  $\phi$-couplings leads to
equations of motion that can be obtained by varying the well-known
D-brane potential:
\be
{\delta\over\delta\phi_m}\frac{\al}{2}\makebox{Tr}(
\phi_{ij}\phi_{ji})=\al [\phi_i,\phi_{im}]=0\, ,
\en
where
\be
\phi_{ij}\equiv [\phi_i,\phi_j]\, .
\en
\section{Two loops in bosonic string theory}
With the sigma model reformulated by using the boundary fermions, the
analysis of higher loop orders becomes much simpler
than in the traditional approach. All diagrams are QED-like, with path-ordered
interaction vertices. In particular, the two-loop computation can be
performed explicitly. There are two diagrams, shown on Fig.3,
which contribute to the two-loop anomalous dimensions.
\begin{figure}
\[
\psannotate{\psboxto(0cm;4cm){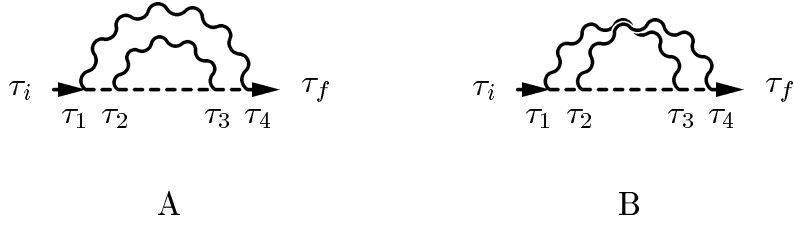}}{}
 \]\vskip -1cm
\caption{\em Two-loop corrections to the boundary fermion propagator.}
\vskip 5mm
\end{figure}
They give, respectively,
\begin{eqnarray}
\Delta^{(3A)}_{ab} &=&
(\phi_i\phi_j\phi_j\phi_i)_{ab}\;{\cal P}\!
\int_{\tau_i}^{\tau_f}
d\tau_1 d\tau_2d\tau_3 d\tau_4
D(\tau_1-\tau_4)D(\tau_2-\tau_3)\, ,\\
\Delta^{(3B)}_{ab} &=&
(\phi_i\phi_j\phi_i\phi_j)_{ab}\;{\cal P}\!
\int_{\tau_i}^{\tau_f}
d\tau_1 d\tau_2d\tau_3 d\tau_4
D(\tau_1-\tau_3)D(\tau_2-\tau_4)\, .
\end{eqnarray}
\begin{figure}
\[
\psannotate{\psboxto(0cm;5cm){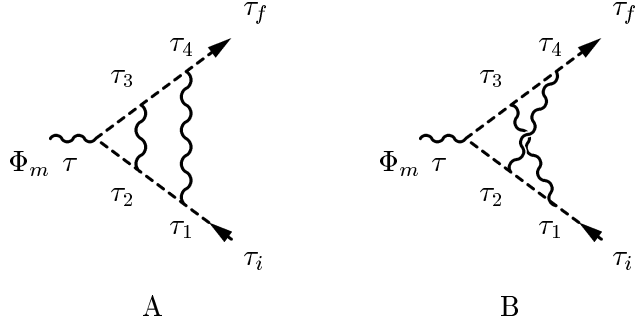}}{}
 \]\vskip -1cm
\caption{\em Two-loop corrections to the boundary fermion-string vertex.}
\vskip 5mm
\end{figure}
The divergent parts of the above integrals can be extracted by elementary
integrations:
\begin{eqnarray}
{\cal P}\!
\int_{\tau_i}^{\tau_f}
d\tau_1 d\tau_2d\tau_3 d\tau_4
D(\tau_1-\tau_4)D(\tau_2-\tau_3)&=&(\al)^2(\frac{1}{2}\ln^2\Lambda-2
\ln\Lambda)+\dots\\
{\cal P}\!
\int_{\tau_i}^{\tau_f}
d\tau_1 d\tau_2d\tau_3 d\tau_4
D(\tau_1-\tau_3)D(\tau_2-\tau_4)&=&(\al)^2 (-\ln^2\Lambda+2\ln\Lambda)+\dots
\end{eqnarray}
In the above expressions, we neglected finite terms as well as
$\ln\Lambda$ terms with ``kinematic'' coefficients
$\ln\sin^{2}[{\scriptstyle{ (\tau_i-\tau_f)\over
2}}]$ which are cancelled by one-loop counterterms.
The two-loop anomalous dimensions can be read out
from the single $\ln\Lambda$ divergences\footnote{
As a consistency check, it is easy to verify that the coefficients
of $\ln^2\Lambda$ terms agree with the renormalization group equations.}
which, together with Eq.(\ref{one}), yield
\be \gamma=  \al\phi_{i}\phi_{i} +(\al)^2\phi_{ij}\phi_{ij} 
+O[(\al)^3].\label{gamma}
\en

The two-loop beta function can be computed in a similar way.
There are only two diagrams, shown on Fig.4, that contribute
single logarithms $\ln\Lambda$ to the vertex function
$\Phi_m(\tau_i,\tau,\tau_f)$. They yield, respectively,
\begin{eqnarray}
\Phi_m^{(4A)} &=&
\phi_i\phi_j\phi_m\phi_j\phi_i\;{\cal P}\!
\int_{\tau_i}^{\tau}\!\!
d\tau_1 d\tau_2
\int_{\tau}^{\tau_f}\!\!d\tau_3 d\tau_4
D(\tau_1-\tau_4)D(\tau_2-\tau_3)\, ,\\
\Phi_m^{(4B)} &=&
\phi_i\phi_j\phi_m\phi_i\phi_j\;{\cal P}\!
\int_{\tau_i}^{\tau}\!\!
d\tau_1 d\tau_2
\int_{\tau}^{\tau_f}\!\!d\tau_3 d\tau_4
D(\tau_1-\tau_3)D(\tau_2-\tau_4)\, .
\end{eqnarray}
For completeness, we list below the results of integrations:
\begin{eqnarray}
{\cal P}\!
\int_{\tau_i}^{\tau}\!\!
d\tau_1 d\tau_2
\int_{\tau}^{\tau_f}\!\!d\tau_3 d\tau_4
D(\tau_1-\tau_4)D(\tau_2-\tau_3)
&=&(\al)^2(\frac{1}{2}\ln^2\Lambda+2\ln\Lambda)+\dots,\\
{\cal P}\!\int_{\tau_i}^{\tau}\!\!
d\tau_1 d\tau_2
\int_{\tau}^{\tau_f}\!\!d\tau_3 d\tau_4
D(\tau_1-\tau_3)D(\tau_2-\tau_4)&=&
-2(\al)^2\ln\Lambda+\dots
\end{eqnarray}
After combining these results with the two-loop anomalous
dimension of Eq.(\ref{gamma}), 
we obtain the following two-loop contribution
to the beta function: \be
\beta_m^{(2)}=(\al)^2[\phi_{ij},[\phi_{ij},\phi_m]]\, . \label{beta}\en
Unlike in the one-loop case, the non-renormalization condition,
$\beta_m^{(2)}=0$, is not integrable, {\em i.e}.\ it cannot be
obtained by varying a gauge-invariant potential. However, by using
Jacobi identity, Eq.(\ref{beta}) can be rewritten as:\be
\beta_m^{(2)}=(\al)^2\left\{ {\delta\over\delta\phi_m}\frac{2}{3}
\makebox{Tr}(
\phi_{ij}\phi_{jk}\phi_{ki})+2[\phi_{im},[\phi_j,\phi_{ji}]]\right\}\, .
\label{fcube}\en
Since the second term involves the one-loop beta function $\beta_i^{(1)}$
of Eq.(\ref{phi}), it is of higher order $O[(\al)^3]$.
In this way, we obtain the D-brane potential\footnote{The overall normalization
is not fixed by the sigma model -- it requires introducing the  brane
tension factor $T_p$.}
\be V(\phi)= \frac{\al}{2}\makebox{Tr}(\phi_{ij}\phi_{ji})
+\frac{2(\al)^2}{3}\makebox{Tr}
(\phi_{ij}\phi_{jk}\phi_{ki})   +O[(\al)^3]\, .\label{pot} \en

Higher loop computations are more difficult, mostly due
to integrals associated to multiple insertions of the interaction vertices.
In particular, the cutoff regularization leads to complicated
integrals and should be replaced
by a more efficient procedure like dimensional regularization.

The scalar potential of Eq.(\ref{pot}) is related by T-duality to
the action for the gauge fields. Indeed, by replacing
$\phi_{ij}\leftrightarrow iF_{\mu\nu}$ and
$[\phi_k,\phi_{ij}]\leftrightarrow D_{\rho}F_{\mu\nu}$, one
obtains the gauge kinetic terms and the  $F^3$ term \cite{ss}. By
a similar transformation on Eq.(\ref{beta}) one recovers the
two-loop beta function computed previously in \cite{dorn,perry}.
More recently, Myers \cite{myers} proposed a complete form of the
scalar potential in superstring theory, to all orders in $\al$, by
relating it via T-duality to the Tseytlin's form of NBI action
\cite{sstrace}. If the approach developed in this work could be
extended to higher loops, it would be useful not only for
computing scalar potentials but also for understanding the
structure of NBI action. In the following Sections, we discuss
superstrings and examine some aspects of Myers' proposal.
\section{Superstrings}
The superstring sigma model action is
\be
S={1\over 2\pi\al}\int_{\Sigma}d^2z\,(\partial X^i\bar{\partial}X^i
+{\al\over 2}\Psi^i\bar{\partial}\Psi^i+{\al\over 2}
\bar\Psi^i\partial\bar\Psi^i)
+S_{\partial\Sigma}(\phi),\label{super}\en
where the world-sheet $\Sigma$ is parameterized
by $z=re^{i\tau}$ and $\bar{z}=re^{-i\tau}$. Here,
$\Psi^i$ and $\bar\Psi^i$ are the world-sheet fermions.
The supersymmetrized Wilson loop corresponds to the interaction term:
\be
S_I=\int_{0}^{2\pi}\!\!d\tau\,[i\partial_rX^i
\lambda^{\dagger}_a(\phi_i)^{ab}\lambda_b\,+\al\psi^i\psi^j
\lambda^{\dagger}_a(\phi_{ij})^{ab}\lambda_b]\, ,\label{siterm} \en
where $\psi^i$ is the restriction of $\Psi^i$ to the boundary, subject
to the Dirichlet condition,\footnote{The  factors of $\sqrt{iz}$ are due
to the $z\rightarrow\tau$ change of variables.}
\be
\sqrt{iz}\Psi^i|_{\partial\Sigma}=-\sqrt{-i\bar{z}}
\bar\Psi^i|_{\partial\Sigma}=i\psi^i(\tau).\en

In the presence of two-form NS background field $B_{ij}(X)$ 
with non-vanishing field strength \be 
H_{ijk}=\partial_i B_{jk}+\partial_k B_{ij}+\partial_j B_{ki}\, ,\label{h}\en
the sigma model contains also the interaction terms which, up
to the linear order in $B_{ij}(X)$, take the form:\footnote{Note that as a
result of analytic continuation to the Euclidean world-sheet this part 
of the action
becomes purely imaginary.} \be
S_B={1\over 2\pi\al}\int_{\Sigma}d^2z\,[B_{ij}(X)\partial X^i\bar{\partial}X^j
+i{\al\over 4}
H_{ijk}\partial X^i\bar\Psi^j\bar\Psi^k
-i{\al\over 4}H_{ijk}\bar{\partial}X^i\Psi^j\Psi^k].\label{sb}\en

In superstring theory, anomalous dimensions and beta functions 
receive additional contributions from the
world-sheet fermions. Their  propagators are \cite{kt}, respectively,
\be \langle\Psi^i(z)\Psi^j(w)\rangle={\delta^{ij}\over z-w},\qquad
\langle\Psi^i(z)\bar\Psi^j(\bar{w})\rangle={i\delta^{ij}\over 1-z\bar{w}}.
\label{fprop}\en
It is also convenient to rewrite the boundary interaction term (\ref{siterm})
as a complex contour integral:
\be
S_I=\oint_{\partial\Sigma}\! du\,[\partial X^i
\lambda^{\dagger}_a(\phi_i)^{ab}\lambda_b\,-\frac{\al}{2}\Psi^i\Psi^j
\lambda^{\dagger}_a(\phi_{ij})^{ab}\lambda_b]-c.c.\label{siz} \en
\begin{figure}
\[
\psannotate{\psboxto(0cm;4cm){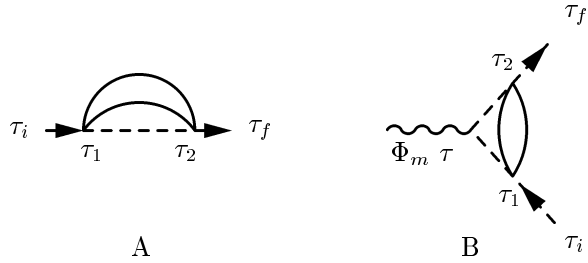}}{}
 \]\vskip -1cm
\caption{\em In superstring theory, starting at the two-loop level, 
world-sheet fermions give rise to
additional contributions to the anomalous dimensions {\rm (A)} and
beta function {\rm (B)}.
World-sheet fermions are represented by solid lines.}
\end{figure}

{}When $B_{ij}=0$, the world-sheet fermions
first contribute at the two-loop level. The relevant diagrams
are shown on Fig.5.
It is easy to see that the diagram 5A cancels the  $O[(\al)^2]$ bosonic
correction to the anomalous dimension (\ref{gamma}) 
while the diagram 5B cancels  the two-loop correction (\ref{beta}) to the
beta function.
As a result, the supersymmetric D-brane potential
does not receive $O[(\al)^2]$ corrections. This
reflects the absence of $F^3$ terms in the supersymmetric NBI action.
\section{NS Fluxes}
As originally pointed out by Myers \cite{myers}, and later confirmed by 
explicit computations of scattering amplitudes \cite{gm}, 
the world-volume potentials involve also couplings of scalars 
to NS fluxes.\footnote{These ``dielectric" couplings involve also
RR fluxes which are beyond the scope of our sigma model considerations.} 
We will first explain  how these couplings
emerge from the sigma model formalism.\footnote{Since we do not discuss
here the derivatives of fluxes,
we will consider the case of constant $H_{ijk}$.}

{}Following the background field method, we introduce a harmonic
background $\widetilde{X}^i$ and expand the interaction term (\ref{sb}):
\be
\widetilde{S}_B={1\over 4\pi\al}\int_{\Sigma}d^2z\,[
\partial \widetilde{X}^iH_{ijk}(\bar{\partial}X^jX^k)
+i\frac{\al}{2}
\partial \widetilde{X}^iH_{ijk}\bar\Psi^j\bar\Psi^k]-c.c. +\dots\label{sbb}\en
Quantum fluctuations can propagate from the
interior of the disk to the boundary  with the (Dirichlet) propagators
\cite{kt}
\be\langle X^i(z)X^j(w)\rangle
=\frac{\al}{2}\delta^{ij}
\Big\{-\ln[(z-w)(\bar{z}-\bar{w})]+\ln[(1-z\bar{w})(1-\bar{z}w)]\Big\}\en
and the fermionic propagators (\ref{fprop}).
Subsequently,  they can 
couple at the boundary to the $\lambda$-fermions  via the interaction 
(\ref{siz}). 
Furthermore, since $\partial \widetilde{X}$ is holomorphic,
we can use Cauchy's formula to express it in terms of the 
boundary data.
After integrating over the interaction position $z$, we obtain the correlation
functions
\begin{eqnarray} \langle \partial X^j(u_1)\partial
X^k(u_2)\rangle_H &=&{i\over 16\pi}\al\oint\! du\,
\partial\widetilde{X}^i(u)\frac{H_{ijk}}{(u_1-u_2)^2}
\ln\left({u_2-u\over u_1-u}
\right),\label{bbb}\\ \langle\Psi^j({u}_1)
\Psi^k({u}_2)\rangle_H 
&=&{1\over 8\pi}\oint\! du\,
\partial\widetilde{X}^i(u)\frac{H_{ijk}}{{u}_1-{u}_2}
\ln\left({u_2-u\over u_1-u}\right)\, ,\label{bff}\end{eqnarray}
where  $u_1$ and $u_2$ are located on the unit 
circle and the integrations extend over the whole circle.\footnote{More
precisely, the $u$-circle should be considered as a limit
$r\to1^+$. This will be relevant later when applying the residue
theorem.}
These correlation functions can be inserted into the Feynman dia\-grams as
shown on Fig.6 and generate ultraviolet divergent contributions
already at one loop. 
\begin{figure}
\[
\psannotate{\psboxto(0cm;4cm){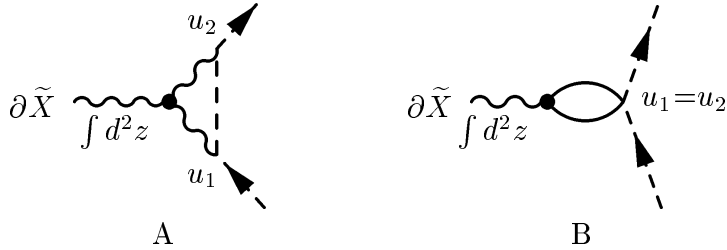}}{}
 \]\vskip -1cm
\caption{\em The one-loop boundary fermion-string interactions involving the 
correlation functions of Eqs.{\rm (\ref{bbb})$\to$(A)} and {\rm 
(\ref{bff})$\to$(B)}. The dotted vertices represent the H-dependent couplings
generated by {\rm (\ref{sbb})}.}
\vskip 5mm
\end{figure}

The diagram of Fig.6A generates the following interaction vertex between
the boundary fermions and the background:
\be I^{(6A)}(u_i,u_f)=-{\al\over 8\pi}H_{mjk}\phi_{jk} 
\oint\! du\,\partial\widetilde{X}^m(u)\int_{u_i}^{u_f}\!\! du_2\int_{u_2}^{u_f}
\frac{du_1}{(u_1-u_2)^2}\ln\left({u_2-u\over u_1-u}\right)
\en
The $u_1$-integral has an ultraviolet divergence at $u_1=u_2$ which can
be regulated by a short-distance cutoff $\Lambda$, 
so that $|u_1-u_2|>\sqrt{\Lambda}$.\footnote{This corresponds to $\sin^2
[{(\tau_1-\tau_2)\over 2}]>\Lambda$, as  previously.}
By integrating over $u_1$, we obtain
\be I^{(6A)}(u_i,u_f)={\al\over 16\pi}H_{mjk}\phi_{jk}\ln\Lambda 
\int_{u_i}^{u_f}\!\! du_2
\oint\!\frac{\partial\widetilde{X}^m(u)}{u-u_2}du +\makebox{finite,} 
\en
which, after using the residue theorem, yields
\be\Phi_m^{(6A)}= {i\al\over 8}H_{mjk}\phi_{jk}\ln\Lambda+
\makebox{finite.}\en
A similar analysis of  diagram 6B shows that it is finite.
We conclude that in the presence of a NS three-form background, the one loop
beta function receives an additional contribution,
\be\beta^{H(1)}_m={i\al\over 4}H_{mjk}\phi_{jk}\, \label{beh},\en
the same for bosonic strings as for superstrings. The corresponding potential
term whose variation ensures $\beta_m=0$ is given by
\be V_H(\phi)=i\frac{\al}{6}
H_{ijk}\makebox{Tr}(\phi_i\phi_j\phi_k)\, .\en
In this way, the ``dielectric'' coupling \cite{myers} appears within the 
framework of sigma model in  superstring theory as well in bosonic string
theory.

It is quite simple to extend the above discussion to the two-loop
order. The only new element is the presence of three-point
vertices that couple quantum fluctuation at the boundary via the
bulk interaction term (\ref{sbb}). After
integrating over the position $z$, we obtain
\begin{eqnarray} \langle \partial X^i(u_1)\partial X^j(u_2)\partial
X^k(u_3)\rangle_H &=&\frac{(\al)^2}{8}
\frac{H_{ijk}}{(u_1-u_2)(u_2-u_3)(u_3-u_1)}\, ,
\label{hbb}\\
\langle\partial
X^i(u_1)\Psi^j({u}_2)\Psi^k({u}_3)\rangle_H
&=&\frac{i\al}{4}\frac{H_{ijk}}{(u_1-u_2)(u_1-u_3)}\, .\label{hff}\end{eqnarray} 
We will consider below the cases of bosonic string and superstring separately.

It is a matter of a straightforward but tedious analysis to show
that in bosonic string theory the only non-vanishing contributions to 
the anomalous dimensions and to the beta function come from 
diagrams shown on Fig.7. \begin{figure}
\[
\psannotate{\psboxto(0cm;3.5cm){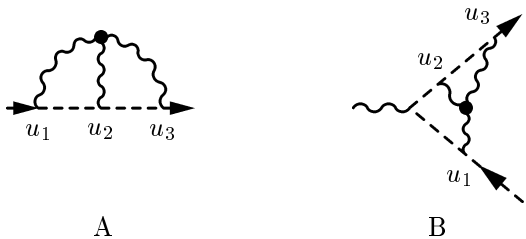}}{}
 \]\vskip -1cm
\caption{\em Two-loop diagrams involving the three-point  vertex
{\rm (\ref{hbb})}.}
\vskip 5mm
\end{figure}\begin{figure}
\[
\psannotate{\psboxto(0cm;3.5cm){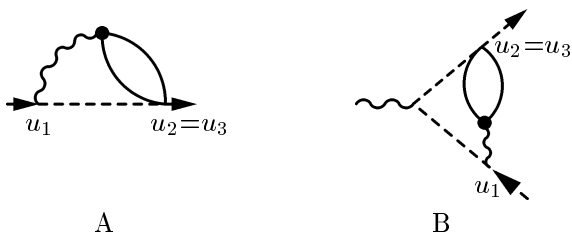}}{}
 \]\vskip -1cm
\caption{\em Two-loop diagrams with the fermionic vertex
{\rm (\ref{hff})}.}
\vskip 5mm
\end{figure}
They yield
\begin{eqnarray} \gamma_H^{(2)}&=&-i(\al)^2H_{ijk}\phi_i\phi_j\phi_k\; ,\\
\beta_{m}^{H(2)}&=&
\frac{i}{2}(\al)^2[\phi_{im},H_{ijk}\phi_{jk}]\; .\label{betah}\end{eqnarray}
In order to read out the corresponding correction to
the world-volume potential one has to combine Eq.(\ref{betah})
with the beta function of  Eq.(\ref{fcube}). It turns out that the additional
contribution combines with the second term of Eq.(\ref{fcube}) to a term of the 
form $2\al [\phi_{im},\beta_i^{(1)}]$, where $\beta_i^{(1)}$ includes now the
$H$-dependent one-loop correction (\ref{beh}). Since such a term is of
order $O[(\al)^3]$, the world-volume
potential does not receive any $H$-dependent corrections at two loops.

In superstring theory there are extra contributions coming from 
diagrams 8A and 8B, see Fig.8. In fact, they cancel against
diagrams 7A and 7B, respectively. Thus in the supersymmetric case,
the world-volume D-brane potential is
\be V(\phi)=\frac{\al}{2}\makebox{Tr}(\phi_{ij}\phi_{ji})+
i\frac{\al}{6}H_{ijk}\makebox{Tr}(\phi_i\phi_j\phi_k)+O[(\al)^3]\, 
\label{vv}.\en
The absence of $O[(\al)^2]$ corrections agrees with Myers' proposal
\cite{myers},  as explained below.

The world-volume  potential proposed by Myers \cite{myers} 
has the form\footnote{We set the D$p$-brane tension $T_p=1$.}
\be V(\phi)= \makebox{STr}\sqrt{\det\{\delta^{i}_{j}+2\pi \al i\phi^{ik}
[G_{kj}(\phi)+B_{kj}(\phi)]\} }\label{vmey}\en
where $G_{kj}(\phi)$ and $B_{kj}(\phi)$ are the transverse metric and
the NS two-form, respectively, whose space-time dependence has been converted
into a functional dependence on scalar matrices by means of a non-abelian Taylor
expansion. In particular,
\be B_{jk}(\phi)=\exp[2\pi\al\phi^i\partial_{X^i}]B_{jk}(X)|_{X^i=0}
=B_{jk}(0)+2\pi\al\phi^i\partial_{i}B_{jk}(0)+\dots\en
The symmetric trace STr prescription \cite{sstrace}
applied to Eq.(\ref{vmey}) amounts to
formally expanding the square root of the determinant followed by
taking the $U(N)$ trace symmetrized
over all possible orderings of the $\phi^i$ and $\phi^{ij}$ matrices.
Note that at the $O(\al)$ order, the potential (\ref{vmey}) does correctly
reproduce Eq.(\ref{vv}) modulo constant factors.
At the next order, the only contribution
involving one derivative of $B_{jk}$, and hence its field strength $H_{ijk}$,
could arise from the $\sqrt{\det}$ term $\phi_{ik}\phi_{jk}
(\phi_m\partial_mB_{ij})$. However, by applying the STr prescription,
one finds that 
this term vanishes after symmetrizing over $\phi_{ik}$ and $\phi_{jk}$.
Hence, the absence of $O[(\al)^2]$ corrections to  D-brane potentials 
lends support to the STr prescription.\footnote{There are two types of 
potential terms, $H_{ijk}$Tr$(\phi_{ij}\phi_k\phi_m\phi_m)$ and
$H_{ijk}$Tr$(\phi_{ij}\phi_m\phi_k\phi_m)$,  allowed 
at that level by gauge and Lorentz invariance, 
so this is indeed a nontrivial check.}
Although many authors have previously
expressed their skepticism as to the validity of this
prescription at higher orders, it certainly holds
up to two loops in the sigma model $\al$ expansion.\\[2ex]
{\bf Acknowledgments}\\
We are grateful to Alexander Barabanschikov, Eric D'Hoker, Zurab Kakushadze,
Boris \linebreak Pioline, Stephan Stieberger and Cumrun Vafa for
illuminating discussions and correspondence.

\newpage
\begin{thebibliography}{99}
\bibitem{revs} For recent reviews, see: J.H. Schwarz, hep-th/0103165;
A.A. Tseytlin, hep-th/9908105.
\bibitem{tsey} A.A. Tseytlin,
\np 276 (1986) 391 [Erratum-ibid.\ B 291 (1987) 876].
\bibitem{dorn} H. Dorn and H.J. Otto, Z. Phys. C 32 (1986) 599.
\bibitem{perry} D. Brecher and M.J. Perry, \np 527 (1998) 121,
hep-th/9801127.
\bibitem{myers} R.C. Myers, JHEP 9912 (1999) 022, hep-th/9910053.
\bibitem{dhok} E. D'Hoker and D.G. Gagn\'e, \np 467 (1996) 272, 
hep-th/9508131.
\bibitem{fermions} S. Samuel, \np 149 (1979) 517;\\
R.A. Brandt, F. Neri, D. Zwanziger, \pr 19 (1979) 1153;\\
J.L. Gervais, A. Neveu, \np 163 (1980) 189;\\
I. Ya. Arefyeva, \pl 93 (1980) 347;\\
H. Dorn, \np 494 (1997) 105; JHEP 9804 (1998) 013.
\bibitem{ss} J. Scherk and J.H. Schwarz, \np 81 (1974) 118.
\bibitem{sstrace} A.A. Tseytlin, \np 501 (1997) 41, hep-th/9701125.
\bibitem{kt}See, {\em e.g}., I.R. Klebanov and L. Thorlacius, \pl 371 (1996) 51,
hep-th/9510200.
\bibitem{gm} M.R. Garousi and R.C. Myers, JHEP 0011 (2000) 032,
hep-th/0010122;\\
A. Fotopoulos, ``Aspects of Duality in Quantum Field Theory and Superstring
Theory,'' Ph.D. Thesis, Northeastern University (2001).
\end{thebibliography}
\end{document}